\documentclass[12pt]{article}
\usepackage{helvet}
\usepackage{authblk}

\usepackage{amssymb,amsbsy,amsfonts,amsmath,amsthm,xspace}
\usepackage{mathrsfs}
\usepackage{courier}
\usepackage[utf8]{inputenc}
\usepackage{tikz}
\usetikzlibrary{calc, matrix, arrows, shapes, positioning, fit, shapes.misc, shapes.geometric, calc, decorations.pathreplacing}
\usepackage{graphicx,multicol,multirow,extarrows,longtable,tabu,booktabs}
\usepackage{newtxtext, newtxmath}
\usepackage{caption}
\usepackage{setspace}
\usepackage{enumitem,dsfont,rotating}
\usepackage{colortbl}

\theoremstyle{plain}

\newtheorem{theorem}{Theorem}[section]

\newtheorem{remark}[theorem]{Remark}
\newtheorem{corollary}[theorem]{Corollary}
\newtheorem{assumption}{Assumption}

\theoremstyle{remark}

\usepackage{subcaption}
\usepackage{epstopdf}
\usepackage{epsfig}
\usepackage{hyperref}
\usepackage{url}
\usepackage[toc,page]{appendix}
\usepackage{float}
\usepackage[authoryear]{natbib}
\usepackage{color}
\usepackage{verbatim}
\usepackage{authblk}
\usepackage{wrapfig}
\usepackage{algorithm2e}
\usepackage[normalem]{ulem}

\usepackage[margin=1in]{geometry}
\newcommand{\bomega}{\boldsymbol{\omega}}

\newcommand{\bSigma}{{\mathbf \Sigma}}

\newcommand{\bgamma}{\boldsymbol{\gamma}}

\newcommand{\btheta} {\boldsymbol{\theta}}

\newcommand{\bA}{{\mathbf A}}

\newcommand{\bD}{{\mathbf D}}

\newcommand{\bX}{{\mathbf X}}
\newcommand{\bY}{{\mathbf Y}}

\newcommand{\ba}{{\mathbf a}}

\newcommand{\bk}{{\mathbf k}}

\newcommand{\bx}{{\mathbf x}}

\newcommand{\bz}{{\mathbf z}}

\def\T{{ \mathrm{\scriptscriptstyle T} }}

\newcommand{\sign}{\text{sign}}

\newcommand{\argmin}{\arg\!\min}
\newcommand{\expect}{{\mathbb{E}}}
\newcommand{\real}{\mathbb{R}}
\def\indicator#1{\mathds{1}\{#1\}}

\usepackage{titlesec}

\titlespacing*{\section}
{0pt} 
{10pt} 
{5pt} 
\titlespacing*{\subsection}
{0pt} 
{6pt} 
{3pt} 

\usepackage{hyperref}
\hypersetup{
  colorlinks   = true, 
  urlcolor     = blue, 
  linkcolor    = blue, 
  citecolor   = blue 
}

\newcommand*\samethanks[1][\value{footnote}]{\footnotemark[#1]}

\title{Modeling and prediction of mutation fitness on protein functionality with structural information using high-dimensional Potts model}

\author[1]{Bingying Dai\thanks{Equal contributions.}}
\author[2]{Yinan Lin\samethanks}
\author[3]{Kejue Jia}
\author[4]{Zhao Ren}
\author[5]{Wen Zhou}

\affil[1]{Department of Statistics, Colorado State University}
\affil[2]{Department of Statistics and Data Science, National University of Singapore}
\affil[3]{Department of Molecular, Cellular and Developmental Biology, Yale University}
\affil[4]{Department of Statistics, University of Pittsburgh}
\affil[5]{Department of Biostatistics, School of Global Public Health, New York University}
\date{}

\begin{document}
\maketitle

\doublespacing
\begin{abstract}
Quantifying the effects of amino acid mutations in proteins presents a significant challenge due to the vast combinations of residue sites and amino acid types, making experimental approaches costly and time-consuming. The Potts model has been used to address this challenge, with parameters capturing evolutionary dependency between residue sites within a protein family. However, existing methods often use the mean-field approximation to reduce computational demands, which lacks provable guarantees and overlooks critical structural information for assessing mutation effects. We propose a new framework for analyzing protein sequences using the Potts model with node-wise high-dimensional multinomial regression. Our method identifies key residue interactions and important amino acids, quantifying mutation effects through evolutionary energy derived from model parameters. It encourages sparsity in both site-wise and amino acid-wise dependencies through element-wise and group sparsity. We have established, for the first time to our knowledge, the $\ell_2$ convergence rate for estimated parameters in the high-dimensional Potts model using sparse group Lasso, matching the existing minimax lower bound for high-dimensional linear models with a sparse group structure, up to a factor depending only on the multinomial nature of the Potts model. This theoretical guarantee enables accurate quantification of estimated energy changes. Additionally, we incorporate structural data into our model by applying penalty weights across site pairs. Our method outperforms others in predicting mutation fitness, as demonstrated by comparisons with high-throughput mutagenesis experiments across 12 protein families.
\end{abstract}

\section{Introduction}
\label{sec:intro}

Evaluating and predicting mutation effects are critical challenges in basic biology and biomedical engineering. Identifying key amino acid alterations underlying organism phenotypes or complex diseases is essential in the ever-expanding catalog of variations in humans and model organisms. While advanced technologies like high-throughput mutational scans have emerged to address these needs \citep{melamed2013deep,lek2016analysis}, assessing protein mutation effects remains challenging due to the high cost and time required. For example, protein mutations can involve hundreds of sites, making it impractical to rely solely on experimental methods.

\begin{figure}[ht]
\centering 
\hspace{-0.5cm}
\includegraphics[width=14.5cm]{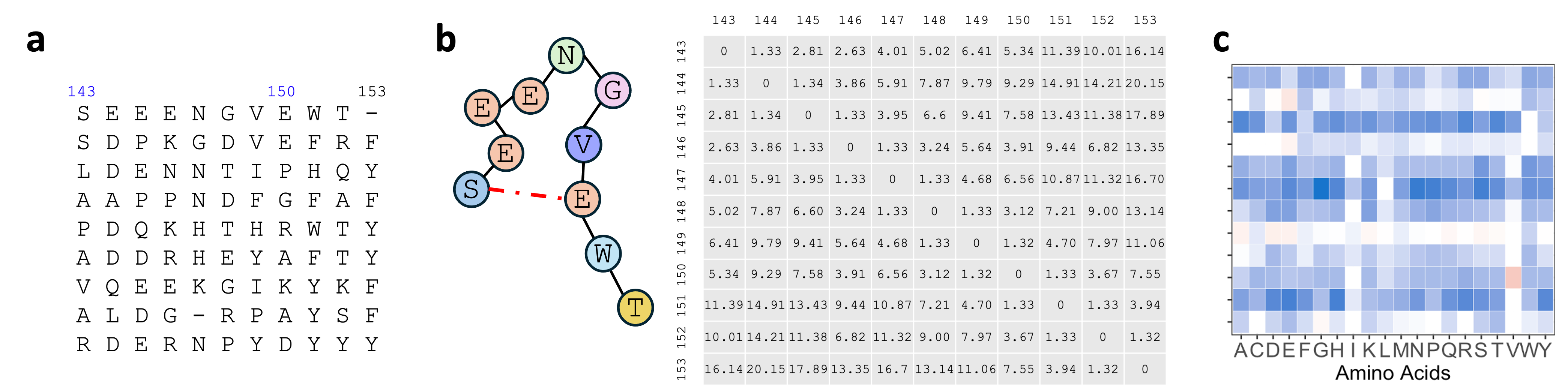}
\caption{(a) MSA data for the DYR family at sites $143$-$153$. (b) Protein structure highlighting the spatial proximity of sites 143 and 150, with site-wise physical distances derived from structural data. The corresponding structure is shown in Figure \ref{fig:DYR}. (c) Integration of MSA and structural data facilitates mutation fitness landscape analysis.}
\label{fig:msa}
\end{figure}

Multiple sequence alignments (MSAs) align residue sites across evolutionarily related protein sequences within the same family and serve as primary input data for studying mutation effects. For example, Figure \ref{fig:msa}(a) showcases a portion of the MSA data for the Dihydrofolate reductase (DYR) protein family, a key enzyme in folate metabolism, responsible for reducing dihydrofolic acid to tetrahydrofolic acid \citep{schnell2004structure}. Each row represents a protein sequence, with twenty distinct letters\footnote{A, C, D, E, F, G, H, I, K, L, M, N, P, Q, R, S, T, V, W, and Y represent amino acids} encoding specific amino acids and dashed lines indicating alignment gaps. MSA data have been widely used in computational methods to assess amino acid substitution impacts, such as SIFT \citep{ng2003sift}, PolyPhen \citep{adzhubei2010method}, and CADD \citep{kircher2014general}. Inspired by the use of maximum entropy, direct coupling analysis uses MSA data to identify essential pair couplings, accounting for sites dependencies \citep{morcos2011direct,levy2017potts}. Similarly, EVmutation (EVM, \cite{hopf2017mutation}) uses MSA data to model mutation fitness landscapes by considering interactions across all protein site pairs. 

The aforementioned computational approaches using MSA data primarily treat it as a collection of fixed alphabets, without quantifying its uncertainty or integrating it with other informative resources, such as protein structural data shown in Figure \ref{fig:msa}(b). In fact, MSA data for each sequence can be viewed as a multivariate categorical random vector. The Potts model \citep{potts1952some}, an extension of the Ising model \citep{ising1925}, naturally model such data, allowing for the investigation of both single-site and pair-site parameters, as formulated in Section \ref{subsec: potts}. In this context, single-site parameters represent site-wise effects, while pair-site parameters, known as direct coupling between sites \citep{morcos2011direct}, capture both site-wise and site-amino acid-wise dependencies. This framework facilitates the calculation of evolutionary statistical energy and provides a coherent assessment of relative energy changes caused by single-site or even multiple-site mutations. Consequently, the Potts model offers a flexible statistical framework for predicting mutation effects and constructing mutation fitness landscapes, as illustrated in Figure \ref{fig:msa}(c).

\subsection{Related works on estimating high-dimensional Potts model}\label{subsec:review} 

Bayesian approaches using Metropolis-Hastings (MH) algorithms are commonly employed for parameter estimation in the Potts model \citep{moller2006efficient, li2017lung}. Due to the computational intractability of the partition function, calculating the MH ratio requires introducing auxiliary variables. This approach leverages the conditional distributions based on the data and additional parameters to eliminate the intractable term. However, several challenges arise, including selecting priors for Potts model parameters, managing burn-in processes, and determining the conditional distributions of auxiliary variables. Sampling these auxiliary variables could be fairly challenging, particularly for high-dimensional cases, requiring either unbiased sampling or accepting Markov chains whose stationary distribution may deviate from the desired posterior \citep{park2018bayesian}. These complexities underscore the need for careful consideration when drawing inferences on the results.

An alternative strategy for parameter estimation in the Potts model is the pseudolikelihood \citep{hopf2017mutation}, which relies on the product of conditional probabilities of one variable given the others \citep{balakrishnan2011learning}. While these methods avoid involving prior knowledge of parameters or auxiliary variables, they are computationally expensive for simultaneous estimation of all parameters. To address this, the Potts-Ising model \citep{razaee2020potts} reduces the dimensionality by constraining pair-site parameters to match the Ising model form. However, this model is tailored for categorical data with a distinctive category, where dependencies depend only on the presence or absence of that category.

As a special case of the Potts model, the Ising model has received significant attention in the literature. Node-wise $\ell_1$-regularized logistic regression \citep{10.1214/09-AOS691} is among the most popular approaches for parameter estimation in the Ising model, with practical algorithms and well-studied theoretical guarantees. It is natural to extend this approach and apply node-wise multinomial regression for parameter estimation in the Potts model, which allows the incorporation of specific penalties. For example, recent work in \cite{tian2023ell_1} investigates the estimation and prediction errors for $\ell_1$-penalized multinomial regression, while other studies explore multinomial regression with structured penalties \citep{tutz2016regularized, nibbering2022multiclass,levy2023generalization}, assuming element-wise or group-wise sparsity. Although \cite{tutz2016regularized} mentions the idea of combining penalties for recapitulating data-specific structures, simultaneous element-wise and group-wise sparsity remains largely underexplored in multinomial regression.
 
\subsection{Our contributions}

In this paper, we propose a high-dimensional Potts model-based statistical framework to analyze MSA data for studying protein mutation effects. We develop parameter estimation for the Potts model using node-wise multinomial regression, which is computationally efficient. Biologically, site-wise dependencies are not expected to be universal to maintain evolutionary flexibility \citep{jones2012psicov,jernigan2021large}, considered as {\it site-wise sparsity}, and strong dependencies between specific amino acids are not densely observed in general, considered as {\it element-wise sparsity}. Thus, we enforce group-wise and element-wise sparsity using the sparse group Lasso penalty \citep{simon2013sparse}, resulting in a node-wise multinomial regression with a double sparse structure at both group and element levels. Additionally, since site-wise dependencies tend to be stronger at spatially closer sites \citep{morcos2011direct, marks2012protein}, we integrate protein structural data with MSA data by deriving group weights in the sparse group Lasso penalty based on spatial distance between sites. Our framework integrates the Potts model, sparse group Lasso, and combined MSA and structural data, providing an efficient statistical approach to predict protein mutation fitness. We apply our method to $12$ protein families to estimate both single-site and pair-site parameters, aiding in the estimation of energy changes due to mutations. As demonstrated in Section \ref{sec: real}, our method achieves much higher correlation with experimentally measured mutation fitness compared to the benchmark EVM \citep{hopf2017mutation}.

Beyond methodological development, we rigorously analyze the theoretical properties of the proposed method. Estimators with sparse group penalties pose challenges due to their non-decomposability \citep{cai2022sparse}. Existing approaches primarily rely on analysis of the Karush–Kuhn–Tucker condition and require additional assumptions, such as incoherence-type condition \citep{cai2022sparse} or strong sparsity conditions \citep{zhang2023high}, to establish estimation error bounds. In this paper, we take a different strategy that avoids these assumptions by deriving a tight upper bound for the stochastic term $\sum_{k=1}^{K} \epsilon_k^{\T}\bX u_k$, induced by the proposed loss function \eqref{eq:Potts_iLS}. Here, $\bX\in \real^{n\times D}$ is the design matrix, $\epsilon_k \in \real^n$ are errors, and $u_k$ is a $D$-dimensional vector. Traditional treatments for $\sum_{k=1}^{K} \epsilon_k^{\T}\bX u_k$, such as $\epsilon_k^{\T}\bX u_k \le \|\bX^{\T}\epsilon_k\|_{\infty}\|u_k\|_1$, may result in sub-optimal convergence rates for estimation \citep{lounici2011oracle}, especially for non-decomposable penalties like sparse group penalties. By modifying the new oracle inequalities for high-dimensional linear models \citep{bellec2018slope}, we derive a finer upper bound for $\sum_{k=1}^{K} \epsilon_k^{\T}\bX u_k$, as detailed after Theorem~\ref{thm:est-err-rates}. This refined bound allows us to establish $\ell_q$ error bounds ($q=1,2$) for the high-dimensional Potts model using sparse group Lasso, matching the minimax lower bound for high-dimensional linear models with sparse group structures, up to a factor dependent on the multinomial nature of the Potts model. This factor highlights fundamental distinctions between multinomial regression and linear or logistic regression, which remain underexplored in existing literature. Furthermore, these error bounds provide consistency guarantees for the estimated energy changes, forming the theoretical foundation for Potts model-based mutation analysis in the literature.

The remainder of this paper is organized as follows. Section \ref{sec:method} provides background on the Potts model and defines evolutionary statistical energy. We then introduce our estimation procedure for the high-dimensional Potts model, employing sparse group Lasso in node-wise multinomial regression. In Section \ref{sec:theory}, we establish the $\ell_1$ and $\ell_2$ convergence rates of the proposed estimators and demonstrate guarantees of the estimated energy changes. Section \ref{sec:structure} discusses the integration of structural information into the Potts model for mutation analysis. Sections \ref{sec: real} and \ref{sec:simulation} evaluate our method through comprehensive real data mutation analyses and simulation studies, both demonstrating the superiority of the proposed method. Finally, Section \ref{sec:dis} concludes with discussions and potential extensions.

\section{Methodology}\label{sec:method}
\subsection{Potts model and evolutionary energy for protein mutations}
\label{subsec: potts}

Consider an MSA data with $n$ protein sequences, each containing $d$ sites. The amino acids at each site are encoded by $K+1$ possible states, starting from $0$, where $K$ generally equals to $20$ and represents the $20$ amino acid types, with $0$ denoting the alignment gap. Denote $[r]=\{1,2,...,r\}$ and $[r]_0=\{0,1,2,...,r\}$ for any positive integer $r$. For each $j \in [d]$, the MSA data at site $j$ can be represented as $\bz_j := (z_{j0},...,z_{jK})^{\T}\in \{0,1\}^{K+1}$, where the binary value $z_{jk}$ indicates whether amino acid $k$ is present at site $j$. It follows that $\sum_{k\in [K]_0} z_{jk} = 1$. An individual protein sequence can then be expressed as $\bz:=(\bz_1^{\T}, \ldots, \bz_d^{\T})^{\T}\in \{0,1\}^{d(K+1)}$. With these notations, the Potts model for a sequence $\mathbf{z}$ with $d$ sites and $K+1$ states is given by
\begin{equation}
    \label{eq:potts model}
    P(\mathbf{z}) = \frac{1}{C} \exp\left\{\sum_{j=1}^d\sum_{k=0}^K\theta_{jk}z_{jk} + \sum_{1\leq j<r\leq d}\sum_{k=0}^K\sum_{l=0}^K \gamma_{jr,kl}z_{jk}z_{rl}\right\},
\end{equation}
where $C$ is the partition function, the parameter $\theta_{jk}$ represents the single-site effect on energy at site $j$ for amino acid $k$, while $\gamma_{jr,kl}$ quantifies the direct coupling between site $j$ with amino acid $k$ and site $r$ with amino acid $l$.

Following the definition in \cite{levy2017potts}, the evolutionary statistical energy (short for {\it energy} hereafter) of a protein sequence $\mathbf{z}$ is defined as \begin{equation} \label{eq:energy function}  E(\mathbf{z}) = \sum_{j=1}^d\sum_{k=0}^K\theta_{jk}z_{jk} + \sum_{1\leq j<r\leq d}\sum_{k=0}^K\sum_{l=0}^K \gamma_{jr,kl}z_{jk}z_{rl}.\end{equation} The overall energy of the sequence is thus the sum of single-site effects and direct couplings between sites, subject to the constraint that only one amino acid type can be assigned to each site. We can represent the energy change of a mutation at single or multiple sites with these parameters. Specifically, consider a mutation at site $j$ that changes the amino acid from the wild-type $a_j$ to another amino acid $k$, while leaving all other sites unchanged. The energy change of this single-site mutation can be calculated as
\begin{equation}
\label{eq:deltaE}
\Delta E_{j,k}=\theta_{jk}-\theta_{ja_j}+\sum_{r\neq j}(\gamma_{jr,ka_r}-\gamma_{jr,a_ja_r}).
\end{equation} This can be naturally extended of a multiple-site mutation. Let $\mathcal{J}$ denote the set of sites where mutations occur, and $\bk_{\mathcal{J}}=(k_j:j\in \mathcal{J})$ denote the amino acids at the mutated sites transitioning from the wild-types. Assuming the cardinality $|\mathcal{J}|>0$, the energy change of such a multiple-site mutation is
\begin{equation}
\label{DeltaE_multi}
    \Delta E_{\mathcal{J},\bk_{\mathcal{J}}} = \sum_{j\in \mathcal{J}} (\theta_{jk_j}-\theta_{ja_j}) +\sum_{j\in \mathcal{J}} \sum_{r\notin \mathcal{J}}(\gamma_{jr,k_ja_r}-\gamma_{jr,a_ja_r})+\sum_{j,j' \in \mathcal{J},j\neq j'}(\gamma_{jj',k_jk_{j'}}-\gamma_{jj',a_ja_{j'}}).
\end{equation}
The energy change corresponds to the log-likelihood ratio between the mutant and wild type, with higher values indicating favorable mutations and lower values for unfavorable ones.

Previous studies \citep{hopf2017mutation} have demonstrated a notable agreement between energy changes and experimental results on mutation fitness. Thus, to analyze mutation fitness using MSA data modeled by the Potts model, it is sufficient to estimate the model parameters. For parameter identifiability, we assign $k=0$ to the wild-type amino acid and set the corresponding parameters $\theta_{j0}$, $\gamma_{jr,0l}$, and $\gamma_{jr,k0}$ to zero. This identifiability adjustment does not affect the definition of $\Delta E_{j,k}$ in \eqref{eq:deltaE}.

\subsection{Node-wise sparse multinomial regression}\label{sec:nodewise-meth}
As mentioned in Section \ref{subsec:review}, directly estimating all parameters in the Potts model \eqref{eq:potts model} can be computationally challenging. Instead, it is more practical to estimate the parameters at each site individually. With a slight abuse of notation, we exclude the wild-type amino acids from $\bz$. For each $j\in[d]$, let $\bz_j = (z_{j1},...,z_{jK})^{\T}\in \{0,1\}^{K}$ represent the amino acid type at site $j$, and let $\bz_{-j}=(\bz_1^{\T}, \ldots, \bz_{j-1}^{\T}, \bz_{j+1}^{\T}, \ldots, \bz_d^{\T})^{\T}\in \{0,1\}^{(d-1)K}$ represent the states at all other sites. Given this setup, when focusing on a single site $j$ while conditioning on the other sites, the conditional probability in the Potts model follows an exponential family distribution that
\begin{equation}
\label{conditional_probability}
    P(\bz_j|\bz_{-j}) =\frac{\prod_{k=1}^K\exp(\theta_{jk}+\sum_{r\neq j}\sum_{l=1}^K\gamma_{jr,kl}z_{rl})^{z_{jk}}}{1 + \sum_{k=1}^K\exp(\theta_{jk}+\sum_{r\neq j}\sum_{l=1}^K\gamma_{jr,kl}z_{rl})},
\end{equation}
Thus, a node-wise multinomial regression can be applied by considering $\bz_j$ for each $j \in [d]$ as the response, representing different amino acid states, and $\bz_{-j}$, the amino acids at all other sites, as covariates. A similar approach is adopted by \cite{cai2019differential} to estimate parameters in an Ising model using node-wise logistic regression.

For each site $j \in [d]$, we first derive the loss function for our node-wise multinomial regression. Since the MSA data at site $j$ is treated as the response, let $\bY_j=(\bY^{(1)}_{j},\bY^{(2)}_j,\ldots,\bY^{(n)}_j)^{\T}$ represent the response matrix of $n$ observed sequences, where $\bY^{(i)}_j=(y^{(i)}_{j1}, \ldots, y^{(i)}_{jK})^{\T}\in \{0,1\}^{K}$ is the $i$th sequence of $\bz_{j}$, and $y^{(i)}_{jk}=1$ indicates that the amino acid at site $j$ in the $i$th sequence is in state $k$. Meanwhile, since the MSA data at all other sites are considered as covariates, let $\bX_{-j} \in \{0,1\}^{n\times(d-1)K}$ represent the design matrix, where the $i$th row is denoted as $\bX_{-j,i}=\bx_{-j}^{(i)}$. Here, $\bx_{-j}^{(i)}$ represents the $i$th observation of $\bz_{-j}$, specifically $\bx_{-j}^{(i)} = (\bx_{1}^{(i)\T},\ldots,\bx_{j-1}^{(i)\T}, \bx_{j+1}^{(i)\T},\ldots,\bx_{d}^{(i)\T})^\T \in \{ 0,1 \}^{(d-1)K}$, where $\bx^{(i)}_r=(\bx^{(i)}_{r1},\ldots,\bx^{(i)}_{rK})^\T \in \{0,1\}^K$ for $r\ne j$. Here, $\bx_{rk}^{(i)}=1$ indicates the amino acid at site $r$ in the $i$th sequence is in state $k$. Given $\bX_{-j}$ and $\bY_{j}$, the negative log-likelihood function is
\begin{equation}
    \label{eq: softmax}
    \ell(\btheta_j, \bgamma_j; \bY_j,\bX_{-j})
     =  \sum_{i=1}^n \left[ \log\left(1+\sum_{l=1}^K \exp\{\theta_{jl}+\bgamma_{j\bullet,k\bullet}^\T \bx^{(i)}_{-j}\}\right) - \sum_{k=1}^K y_{jk}^{(i)}(\theta_{jk}+\bgamma_{j\bullet,k\bullet}^\T \bx^{(i)}_{-j})\right],
\end{equation}
where $\btheta_j=(\theta_{j1},...,\theta_{jK})^\T \in \mathbb{R}^K$ consists of the single-site effects for the $K$ amino acids at site $j$, and  $\bgamma_j=(\bgamma_{j\bullet,1\bullet}^\T,...,\bgamma_{j\bullet,K\bullet}^\T) \in \mathbb{R}^{(d-1)K^2}$ consists of all possible dependencies between the $K$ amino acids at site $j$ and the $K$ amino acids at each of the other $d-1$ sites. For each $k\in [K]$, $\bgamma_{j\bullet,k\bullet}=(\bgamma_{j1,k\bullet}^{\T},\ldots, \bgamma_{j(j-1),k\bullet}^{\T}, \bgamma_{j(j+1),k\bullet}^{\T}, \ldots, \bgamma_{jd,k\bullet}^{\T})^{\T} \in \mathbb{R}^{(d-1)K}$ represents the dependencies between site $j$ with amino acid $k$ and the $K$ amino acids at each of the other $d-1$ sites. In addition, $\bgamma_{jr,k\bullet}=(\gamma_{jr,k1}, \ldots, \gamma_{jr,kK})^{\T} \in \mathbb{R}^{K}$ for $r\ne j$ collects the dependencies between site $j$ with amino acid $k$ and the $K$ amino acids at site $r$.

Given a fixed site $j$ and for each site pair $(j,r)$ with $r\ne j$, the direct coupling 
\begin{equation}
    \bgamma_{j(r)}=(\bgamma_{jr,1\bullet}^{\T}, \ldots, \bgamma_{jr,K\bullet}^{\T})^{\T}\in \mathbb{R}^{K^2}
    \label{eq:gamma-group-structure}
\end{equation}
between the $K$ amino acids at site $j$ and the $K$ amino acids at site $r$ can be viewed as a subgroup within $\bgamma_{j}$. Particularly, sites $j$ and $r$ are independent conditional on other sites if and only if $\bgamma_{j(r)}=\mathbf{0}$. As discussed in Section \ref{sec:intro}, dependencies are not expected for all site pairs, indicating {\it site-wise sparsity}. Also, strong dependencies between specific amino acids, primarily driven by biochemical properties like polarity and charge, are neither expected to be universal, suggesting {\it element-wise sparsity} within amino acid groups. To account for these, we enforce both group-wise and element-wise sparsity in our procedure. This leads to the penalty $h(\bgamma_j)= \lambda_g\sum_{r\neq j} \|\boldsymbol{\bgamma}_{j(r)}\|_2+\lambda \sum_{r\neq j}\|\bgamma_{j(r)}\|_1$, where $\lambda_g, \lambda > 0$ are the tuning parameters for the group Lasso and Lasso penalties, respectively, and $\|\cdot\|_q$ denotes the $\ell_q$-norm of a vector for a positive integer $q$. In this context, the group Lasso penalty \citep{yuan2006model} selects sites with strong dependencies with the site $j$, while the Lasso penalty \citep{tibshirani1996regression} identifies significant dependencies at the amino acid level. Combining these penalties results in the sparse group Lasso penalty \citep{simon2013sparse}, which enables a double sparse structure at both the group and element levels. Thus, for each site $j\in [d]$, we propose the following risk function with sparse group Lasso as the objective:
\begin{equation}
\label{eq:Potts_iLS}
    L(\btheta_j, \bgamma_j) = \ell(\btheta_j, \bgamma_j; \mathbf{Y}_j,\mathbf{X}_{-j}) + h(\bgamma_j),
\end{equation}
where $\ell(\boldsymbol{\gamma}_j,\btheta_j; \mathbf{Y}_j,\mathbf{X}_{-j})$ is specified in \eqref{eq: softmax}. 
With this objective function, our estimator for $(\btheta_j, \bgamma_j)$ of site $j$ is
\begin{equation}
	(\widehat\btheta_j, \widehat\bgamma_j) = \argmin_{\btheta_j, \bgamma_j} L(\btheta_j, \bgamma_j).
	\label{eq:sgLasso-est}
\end{equation}

\begin{remark}\label{re:group_weights}
To further incorporate additional structural information, group-specific weights $w_{jr}$ can be introduced into the group Lasso penalty, resulting in the following modified penalty function
\begin{equation}
\label{eq:penalty_weighted}
    h(\bgamma_j)= \lambda_g \sum_{r\neq j} w_{jr}\|\boldsymbol{\bgamma}_{j(r)}\|_2+\lambda\sum_{r\neq j}\|\bgamma_{j(r)}\|_1.
\end{equation}
These weights can be selected to reflect the expectation that sites in closer proximity are likely to exhibit stronger dependencies, allowing the model to account for underlying spatial structures more effectively. A further discussion is detailed in Section \ref{sec:structure}.
\end{remark}

\subsection{Parameter estimations}\label{sec:est-proc}

For each $j \in [d]$, although the optimization in \eqref{eq:sgLasso-est} does not have a closed-form solution, it can be solved with a coordinate gradient descent algorithm \citep{vincent2014sparse}. The algorithm consists of a sequence of nested loops: the outer one relies on a quadratic approximation of \eqref{eq: softmax} at the current iteration, while the middle and inner loops focus on solving the within-group subproblems as described in \cite{simon2013sparse}. Notably, the estimates $(\widehat{\btheta}_j, \widehat{\bgamma}_j)$ obtained from the node-wise regressions do not inherently guarantee the symmetry of $\gamma_{jr,kl}$ and $\gamma_{rj,lk}$. To enforce symmetry, we post-process the estimates by averaging the two values.

Let the gradient of $\ell(\btheta_j, \bgamma_j; \bY_j,\bX_{-j})$ at $\bgamma_j$ be denoted as $\nabla \ell(\bgamma_j)$ and its Hessian matrix as $\nabla^2 \ell(\bgamma_j)$. Here, $\nabla \ell(\bgamma_j)\in \mathbb{R}^{(d-1)K^2}$ and $\nabla^2 \ell(\bgamma_j)\in \mathbb{R}^{(d-1)K^2 \times (d-1)K^2}$. We use $\nabla \ell(\bgamma_j)_{(r)}$ and $[ \nabla^2 \ell(\bgamma_j) \nabla \ell(\bgamma_j) ]_{(r)}\in \mathbb{R}^{K^2}$, following the same group order defined in \eqref{eq:gamma-group-structure}. Define the coordinate-wise soft thresholding operator $S(\ba,b)$ for a vector $\ba =(a_1,...,a_n)^\T$ and a constant $b$ as $S(\ba,b)=(\sign(a_1)\max\{|a_1|-b,0\},...,\sign(a_n)\max\{|a_n|-b,0\})^\T$. The estimation procedure for fixed $\lambda$ and $\lambda_g$ is outlined in Algorithm \ref{alg:cd}, while $(\lambda, \lambda_g)$ can be tuned via 5-fold cross-validation, as described in Section \ref{sec:simulation-setting}. In Section \ref{sec:nodewise-meth}, we take the MSA data at each site as the response and the data from the remaining sites as covariates. This allows us to obtain $(\widehat\btheta_j, \widehat\bgamma_j)$ for all $j \in [d]$ less costly by leveraging parallel computation.

\RestyleAlgo{ruled}
\begin{algorithm}[ht]
\caption{Parameter estimations for the Potts model \eqref{eq:potts model}}
\KwIn{MSA data, initialization of $\widehat\btheta_j$, $\widehat\bgamma_j$ for $j\in [d]$, and penalty parameters $\{ \lambda,\lambda_g$\}}
For each $j\in [d]$, extract $\{\bX_{-j}^{(i)},\bY_{j}^{(i)}\}_{i=1}^n$, \Repeat{convergence}{ 
    For each $r \in [d]$, $r \neq j$, \Repeat{convergence}{
            Update $\widehat{\btheta}_{j}^{\text{new}}$ by mean centering as suggested by \cite{friedman2010regularization};  \\
            Define the block gradient 
            \[\nabla g_{jr}(\bgamma_j)=\nabla\ell(\widehat\bgamma_j)_{(r)}+[\nabla^2\ell(\widehat\bgamma_j)(\bgamma_j- \widehat \bgamma_j)]_{(r)}\in \mathbb{R}^{K^2}.\] \\
            \textbf{if}{ $\|S(\nabla g_{jr}(\mathbf{0}),\lambda)\|_2 \leq \lambda_g $, \textbf{then }$\widehat{\bgamma}_{jr}^{\text{new}}=\mathbf{0}$.} \\
            \textbf{else}{
             Minimize the objective function over each component in $\widehat{\bgamma}_{jr}^{\text{new}}$ \citep{simon2013sparse}.
            }
        }
    $\widehat{\bgamma}_j = \widehat{\bgamma}_j + t(\widehat{\bgamma}_j-\widehat{\bgamma}_j^{\text{new}})$ and 
    $\widehat{\btheta}_j = \widehat{\btheta}_j + t(\widehat{\btheta}_j-\widehat{\btheta}_j^{\text{new}})$ with step size $t$ from line search.
    }  

\For{$j\in [d],r>j$ and $k,l\in[K]$}{
 Post-process estimates for symmetry: $\Tilde{\gamma}_{jr,kl}= \Tilde{\gamma}_{rj, lk} = \frac{1}{2}(\widehat{\gamma}_{jr,kl} + \widehat{\gamma}_{rj,lk})$.
}
\KwOut{$\widehat{\btheta}_j$, $\Tilde\bgamma_j$} 
\label{alg:cd}
\end{algorithm}

\section{Theoretical Guarantee}\label{sec:theory}
In this section, we derive non-asymptotic $\ell_1$ and $\ell_2$ error bounds for the proposed sparse group Lasso estimator \eqref{eq:sgLasso-est}. Building on these results, we also establish theoretical guarantees for the plug-in estimators of  energy changes. 

\subsection{Preliminaries and assumptions}

As described in Section \ref{sec:nodewise-meth}, for each site $j\in [d]$, the dependency parameter vector $\bgamma_j \in \real^{D}$, where $D=(d-1)K^2$, has a group structure such that elements in $\bgamma_j$ can be divided into $d-1$ groups $\{\bgamma_{j(r)}: r\ne j\}$,
where $\bgamma_{j(r)}$ corresponds to the group (site) $r$ as defined in \eqref{eq:gamma-group-structure}. Moreover, let $\gamma_{jl}$ for $l\in [D]$ represent the $l$th element of $\bgamma_j$. Given two positive integers $s$ and $s_g$ satisfying $s_g \le d-1$ and $s_g \le s \le s_g K^2$, we say $\bgamma_j$ is $(s, s_g)$-sparse if $\|\bgamma_j\|_{0,2} := \sum_{r\ne j} \indicator{\bgamma_{j(r)}\ne 0} \le s_g$ and $\|\bgamma_j\|_{0} := \sum_{r\ne j}\sum_{k,l\in[K]}\indicator{\gamma_{jr,kl}\ne 0} \le s$, where $\|\cdot\|_{0}$ represents the vector $\ell_0$-norm.
Let $\btheta_j^{*}$ and $\bgamma_j^{*}$ represent the true parameter vectors in the Potts model \eqref{eq:potts model} for the site $j$, and define $\bgamma_j^{\circ*}=(\btheta_j^{*\T}, \bgamma_j^{*\T})^{\T}$. We assume for each $j\in [d]$ that the true coefficient vector $\bgamma^{*}_j$ is $(s, s_g)$-sparse. That is, for the Potts model, we have
\[
	\max_{j\in [d]} \|\bgamma_j\|_{0,2} \le s_g
	\quad \text{and}\quad 
	\max_{j\in [d]}\|\bgamma_j\|_{0} \le s.
\]
Since entries in $\btheta_j^{*}$ are often assumed to be non-zero for all $j\in[d]$ to model the MSA data, then $\bgamma_j^{\circ*}$ is $(s^{\circ}, s_g^{\circ})$-sparse with $s^{\circ}=s+K$ and $s_g^{\circ}=s_g+1$ by treating $\btheta_j^{*}$ as an additional group.

Recall that, for each $j\in [d]$ and $i\in [n]$, the $i$th row in the design matrix $\bX_{-j}$ is $\bx_{-j}^{(i)} \in \{ 0,1 \}^{(d-1)K}$. To investigate the theoretical properties of $\widehat{\bgamma}^{\circ}_j=(\widehat\btheta_j^{\T}, \widehat\bgamma_j^{\T})^{\T}$ from \eqref{eq:sgLasso-est}, we propose the following assumptions. 

\begin{assumption}\label{as:eigen}
	For each $j\in [d]$, suppose the rows $\{\bx_{-j}^{(i)}: 1\le i \le n\}$ in the design matrix are independent and identically distributed random vectors with 
$\bSigma = \expect[(1 ~ (\bx_{-j}^{(1)})^{\T})^{\T}  (1 ~ (\bx_{-j}^{(1)})^{\T})]$, where $\bSigma$ satisfies $\lambda_{\min}(\bSigma)\ge c_{\lambda}$ for some $c_{\lambda} >0$.
\end{assumption}

\begin{assumption}\label{as:margin}
	With probability larger than $1-M_n$ where $M_n\to 0$ as $n\to \infty$, there exists an absolute constant $C_0>0$, such that
$\max_{i\in[n], k\in[K], j\in [d]} |\zeta_{jk}^{(i)}| \le C_0$ with $\zeta_{jk}^{(i)} =\theta_{jk}^{*}+\bgamma_{j\bullet,k\bullet}^{*\T} \bx_{-j}^{(i)}$,
which implies that $\min_{i\in[n], k\in[K], j\in [d]}  \exp(\zeta_{jk}^{(i)})\{1+\sum_{l=1}^{K}\exp(\zeta_{jl}^{(i)})\}^{-1}\ge c_*$ and $\min_{i\in[n], j\in [d]}  \{1+\sum_{l=1}^{K}\exp(\zeta_{jl}^{(i)})\}^{-1}\ge c_*$, for some $c_*>0$.
\end{assumption}

Conditions similar to Assumption \ref{as:eigen} regarding the eigenvalues of $\bSigma$ are commonly adopted in high-dimensional statistics \citep{cai2022sparse, zhang2023high, tian2023transfer}. Unlike many studies on high-dimensional regressions, the rows in our design matrix are not sub-Gaussian vectors but instead have bounded entries. This distinction imposes a stronger sparsity requirement compared to sub-Gaussian cases; see the discussions following Theorem \ref{thm:est-err-rates} for more details. Conditions similar to Assumption \ref{as:margin} are also frequently used for the high-dimensional multinomial regression \citep{tian2023ell_1,abramovich2021multiclass} and high-dimensional logistic regression \citep{guo2021inference, ma2022statistical}. The factors $c_{\lambda}$ and $c_*$ play important roles in establishing the convergence rate of an estimator in multinomial regression; see more discussions after Theorem \ref{thm:est-err-rates}. Notably, the Potts model reduces to the Ising model when $K=1$, and similar assumptions on $c_{\lambda},c_*=O(1)$ are often imposed in literature \citep{10.1214/09-AOS691, cai2019differential}.

\subsection{Guarantees on estimated parameters}
For a constant $\eta\in(0, 1/2)$, define $\delta(\lambda)= \exp[-\{(\eta\lambda\sqrt{n})/40\}^2]$ so that $\lambda = 40\eta^{-1}\sqrt{n^{-1}\log(1/\delta(\lambda))}$. Set
\begin{equation}
	\lambda_{\sharp}=40\sigma(\eta n^{1/2})^{-1} \{2(s_0^{\circ})^{-1}\log(4ed/s_g^{\circ})+\log(2eK^2/s_0^{\circ})\},
	\label{eq:def-lambda-sharp}
\end{equation}
where $s_0^{\circ}=s^{\circ}/s_g^{\circ}$ represents the average sparsity per group in the true groups, and $\sigma^2 = \max_{i,k}\text{Var}(y_{k}^{(i)}|\bx^{(i)})$ is the conditional variance of the responses. For non-negative sequences $\{a_n\}$ and $\{b_n\}$, $a_n = o(b_n)$ or $a_n \ll b_n$ means $a_n / b_n \to 0$ as $n \to \infty$, and $a_n = O(b_n)$ or $a_n\lesssim b_n$ means $a_n \le c b_n$ for some absolute constant $c>0$ and sufficiently large  $n$. Recall that $\bgamma_j^*$ are $(s, s_g)$-sparse for each $j$ and $\bgamma_j^{*\circ}$ are $(s^{\circ}, s_g^{\circ})$-sparse with $s^{\circ}=s+K$ and $s_g^{\circ}=s+1$. We have the following guarantees on the proposed estimators. 

\begin{theorem}\label{thm:est-err-rates}
	Suppose Assumptions \ref{as:eigen} and \ref{as:margin} hold, and $s^{\circ}\ll c_*^2 \sqrt{n/\log(dK)}$. 
	Let $\delta_0\in (0,1)$ satisfies $\log(1/\delta_0)\{s^{\circ}\log(1/\delta(\lambda_{\sharp}))\}^{-1} = O(1)$.
	For each site $j\in [d]$, there exists an absolute constant $C>0$, such that with probability at least $1-C((d-1)K+1)^{-1}-M_n-\delta_0$, the estimator $(\widehat{\btheta}_j, \widehat{\bgamma}_j)$ from \eqref{eq:sgLasso-est} with $\lambda = 2\lambda_{\sharp}$ and $\lambda_{g} = \sqrt{s/s_g} \lambda$
	satisfies for $q=1,2$\begin{equation}
		(\|\widehat{\btheta}_j-\btheta^{*}_j\|_q^q + \|\widehat{\bgamma}_j-\bgamma^{*}_j\|_q^q)^{1/q}
		\le
		C R_{K} B_n,
		\label{eq:error-bound-thm}
	\end{equation}
where $R_{K}= \sigma/(c_{\lambda} c_*^2)$ and $B_n = (s^{\circ})^{1/q}n^{-1/2}\{2(s_0^{\circ})^{-1}\log(4ed/s_g^{\circ})+\log(2eK^2/s_0^{\circ})\}^{1/2}.$
\end{theorem}    

The condition $s^{\circ}\ll c_*^2\sqrt{n/\log(dK)}$ is necessary for two main reasons. First, the factor $c_*^2$ arises from the multinomial nature of the Potts model, particularly in the quadratic approximation analysis for the likelihood function. Second, while sub-Gaussian designs for high-dimensional regression typically require $s^{\circ}\ll c_*^2 n/\log(dK)$, our design matrix, with rows $\{\bx_{-j}^{(i)}, 1\le i \le n\}$, has bounded elements and falls outside the sub-Gaussian framework. In such cases, stricter sparsity requirement for $s^{\circ}$ is often required, as noted in Theorem 2.4 of \cite{van2014asymptotically}. 

In Theorem \ref{thm:est-err-rates}, $\lambda$ and $\lambda_g$ depend on $(s, s_g)$, while the convergence rate is determined by $(s^{\circ}, s_g^{\circ})$. This arises because our estimation procedure penalizes only the dependency vector $\bgamma_j$. As the true dependency vector $\bgamma_j^*$ is $(s, s_g)$-sparse, it is intuitive that the tuning parameters are related to $(s, s_g)$. Since the single-site effects vector $\btheta_j$ is not penalized and $\bgamma_j^{\circ*}$ is $(s^{\circ}, s_g^{\circ})$-sparse, leading to a convergence rate dependent on $(s^{\circ}, s_g^{\circ})$.

The convergence rate in \eqref{eq:error-bound-thm} is $B_n$ when $R_K$ is an absolute constant, which holds if $K$ is fixed, including the important Ising model with $K=1$. In this case, the convergence rate in Theorem \ref{thm:est-err-rates} matches the minimax optimal rate for linear models. By taking $\delta_0= \exp[-C_2\{2s_g^{\circ}\log(4ed/s_g^{\circ})+s^{\circ}\log(2eK^2/s_0^{\circ})\}]$, we have $\delta_0=o(1)$ as long as $d$ diverges. The condition $\log(1/\delta_0)\{s^{\circ}\log(1/\delta(\lambda_{\sharp}))\}^{-1} = O(1)$ is then satisfied with this $\delta_0$. As a result, with probability approaching one as $n\to \infty$, we have
\begin{equation}
	(\|\widehat{\btheta}_j-\btheta^{*}_j\|_2^2 +  \|\widehat{\bgamma}_j-\bgamma^{*}_j\|_2^2)^{1/2}
	\lesssim \frac{1}{\sqrt{n}}
	\left\{
	2s_g^{\circ}\log\left(\frac{4ed}{s_g^{\circ}}\right)+
	s^{\circ}\log\left(\frac{2es_g^{\circ} K^2}{s^{\circ}}\right)
	\right\}^{1/2}.
	\label{eq:l2-error}
\end{equation}
Compared to the $\ell_2$ error bounds in \cite{cai2022sparse} and \cite{zhang2023high} for linear models with sparse group structures, our error bound in \eqref{eq:l2-error} sharpens their results by reducing logarithmic factors. Specifically, the second term in \eqref{eq:l2-error} from previous works is at least of order $s^{\circ}\log(eK^2)$, which is strictly larger than our bound. Moreover, our error bound tightly matches the minimax lower bound, including the logarithmic factors, derived in \cite{cai2022sparse} for linear models. Our analysis leverages sharp oracle inequalities recently developed in \cite{bellec2018slope} and \cite{li2023sharp} for high-dimensional linear models with $\ell_1$- and sparse group Lasso penalties. Unlike standard Lasso-type analysis \citep{wainwright2019high}, the new oracle inequalities are based on a refined analysis of the concentration of the sum of the leading entries in the non-increasing rearrangement of the magnitudes of the products of covariates and error terms. However, adapting this analysis to the Potts model requires substantial modifications; see, for example, Lemma B1 of the Supplementary Material for details.

In \eqref{eq:error-bound-thm}, $R_K$ may depend on $K$ and is determined by three factors: the smallest eigenvalue of $\bSigma$ for the covariates, governed by $c_{\lambda}$; the minimum success probability level $c_*$; and the conditional variance $\sigma^2$ of the response. The product $c_{\lambda}c_*^2$ provides a lower bound for the smallest eigenvalue of the Hessian matrix of the log-likelihood function, which is typically assumed to be $O(1)$ in linear or logistic regression. In logistic regression, $\sigma^2=O(1)$ under a common boundedness condition similar to Assumption \ref{as:margin}. In multinomial regression, even with Assumption \ref{as:margin}, the variance level $\sigma^2$ is of order $O(1/K)$, differing fundamentally from logistic regression. While $c_{\lambda}=O(1)$ is a natural assumption, the smallest eigenvalue of the Hessian matrix for the multinomial log-likelihood is tightly bounded below by $c_*^2=O(1/K^2)$, as shown in Lemma B5 of the Supplementary Material. This suggests that the multinomial regression likelihood exhibits singularities as $K$ diverges. This characterization of $c_*^2$ is a substantial distinction between multinomial and logistic regression that remains underexplored in the literature. Notably, in \cite{tian2023ell_1}, the authors studied $\ell_1$-penalized multinomial regression under similar assumptions as ours, with $c_{\lambda}=O(1)$. However, their $\ell_2$ error bound involves $K^{5/2}$ factor, significantly larger than our result with $K^{3/2}$. Finally, the Potts model shares all the intrinsic features of multinomial regression but admits $c_{\lambda}=O(1/K)$ due to its unique covariate structure. This can be intuitively understood by considering a case where all sites in $\bz_{-j}$ are independent for some site $j$. Here, $c_{\lambda}=\min_{0\le k\le K} p_{jk}=O(c_*)=O(1/K)$, where $p_{jk}=P(z_{jk}=1)$. 

Theorem \ref{thm:est-err-rates} shows that the proposed estimator $(\widehat{\btheta}_j, \widehat{\bgamma}_j)$ from \eqref{eq:sgLasso-est} achieves an improved error bound when the parameter vector is simultaneously element-wise and group-wise sparse. Recall that $\bgamma_j^{\circ*}=(\btheta_j^{*}, \bgamma_j^{*})$ has length $D^{\circ}=K+(d-1)K^2$, with $d$ groups and is $(s^{\circ}, s_g^{\circ})$-sparse. Using only an $\ell_1$ penalty $\lambda\|\bgamma\|_1$ in \eqref{eq:sgLasso-est} yields an $\ell_2$ estimation error of order $\sqrt{s^{\circ}n^{-1}\log(D^{\circ}/s^{\circ})}$ \citep{bellec2018slope}, larger than the order in \eqref{eq:l2-error} when $s_g^{\circ}/d=o(1)$ and $s_g^{\circ}/s^{\circ}=o(1)$. Similarly, using only a group penalty $\lambda_g\sum_{r\ne j}\|\gamma_{j(r)}\|_2$ results in an $\ell_2$ error bound of order $\sqrt{s_g^{\circ}n^{-1}(\log d + K^2)}$ \citep{lounici2011oracle}, which exceeds the order in \eqref{eq:l2-error} when $\log d/K^2=o(1)$ and $s^{\circ}/s_g^{\circ}=o(K^2/\log K^2)$. Thus, the proposed estimator achieves a tighter $\ell_2$ error bound compared to 
estimators using only $\ell_1$- or group penalties, when the true underlying parameter vector is both element-wise and group-wise sparse.

\subsection{Consistency on estimated energy for mutation fitness}

From \eqref{eq:error-bound-thm}, we also have the $\ell_1$ error bound of the proposed estimator:
\[
\|\widehat{\btheta}_j-\btheta^{*}_j\|_1 + \|\widehat{\bgamma}_j-\bgamma^{*}_j\|_1
\lesssim
R_K\sqrt{\frac{s^{\circ}}{n}}
\left\{2s_g^{\circ}\log\left(\frac{4ed}{s_g^{\circ}}\right)
	+
	s^{\circ}\log\left(\frac{2es_g^{\circ} K^2}{s^{\circ}}\right)\right\}^{1/2},
\]
which immediately implies the consistency for $\widehat{\Delta E}_{j,k}$, the plug-in estimated relative energy change using $(\widehat{\btheta}_j, \widehat{\bgamma}_j)$ from \eqref{eq:sgLasso-est} for $\Delta E_{j,k}$ defined in \eqref{eq:deltaE}, as summarized below. 
\begin{corollary}
	Under the assumptions of Theorem \ref{thm:est-err-rates}, for each site $j\in [d]$ {and amino acid type $k\in [K]$}, with the same probability therein, we have 
	$|\widehat{\Delta E}_{j,k} - \Delta E_{j,k}|\lesssim R_{K} B_n$, where $R_K$ and $B_n$ are defined in Theorem \ref{thm:est-err-rates}.
\end{corollary}

Building on the results for each site in the Potts model, we can straightforwardly establish global results for the entire sequence. To enforce symmetry in the estimator $(\widehat{\btheta}_j, \widehat{\bgamma}_j)$ from \eqref{eq:sgLasso-est} for site $j$, we average the estimates in Section \ref{sec:est-proc}, yielding the symmetrized estimator $\Tilde{\bgamma}_j$. Specifically, $\Tilde{\gamma}_{jr,kl}= \Tilde{\gamma}_{rj, lk} = (\widehat{\gamma}_{jr,kl} + \widehat{\gamma}_{rj,lk})/2$ for each $r\ne j$. 
Below, we summarize the global results for $\{(\widehat{\btheta}_j, \Tilde{\bgamma}_j): j\in [d]\}$, based on which the theoretical guarantees for the plug-in estimator of the energy changes for multiple-site mutations can be readily established. 

\begin{corollary}
	Under the assumptions of Theorem \ref{thm:est-err-rates}, with the same probability and $R_K, B_n$ specified therein, we have
	\[
		\sum_{j=1}^d (\|\widehat{\btheta}_j-\btheta^{*}_j\|_2^2 + \|\tilde{\bgamma}_j-\bgamma^{*}_j\|_2^2)
		\lesssim
		d R_{K}^2 B_n^2
		~~
		\text{and}
		~~ 
		\sum_{j=1}^d (\|\widehat{\btheta}_j-\btheta^{*}_j\|_1 + \|\tilde{\bgamma}_j-\bgamma^{*}_j\|_1)
		\lesssim
		d R_{K} B_n.
	\]
\end{corollary}

\section{Integration of Structural Information via Group Weights}\label{sec:structure}

The sparse group Lasso penalty ensures the desired site-wise and element-wise sparsity in our method but may not fully leverage the dependencies between spatially proximal sites \citep{marks2012protein}. To address this, we first retrieve one representative 3-dimensional protein structure (Alphafold prediction) for each family from the UniProt database \citep{gkaa1100}. For each site pair $(j,r)$, spatial proximity is calculated as the Euclidean distance (in $\mathring{A}$) between the 3-dimensional coordinates of their respective alpha-carbon atoms, denoted as $D_{jr}$.

To integrate this structural information into our method, we then introduce group weights determined by physical distances within the protein's structure. That is, for each site $j\in[d]$, we define the group weight $w_{jr}$ in \eqref{eq:penalty_weighted} as
\begin{equation}
    w_{jr}= (\sqrt{d_r/n}+\sqrt{2\log(d-1)/n})  \mathcal{K}(D_{jr}),
    \label{eq:def-group-weight}
\end{equation}
where $d_r$ denotes the group size of $\bgamma_{j(r)}$, specifically $K^2$ as defined in \eqref{eq:gamma-group-structure}, and $\mathcal{K}(D_{jr})$ is a function that incorporates structural information into the weights.

\begin{figure}[ht]
\centering 
\hspace{-0.6cm}
\includegraphics[scale=0.425]{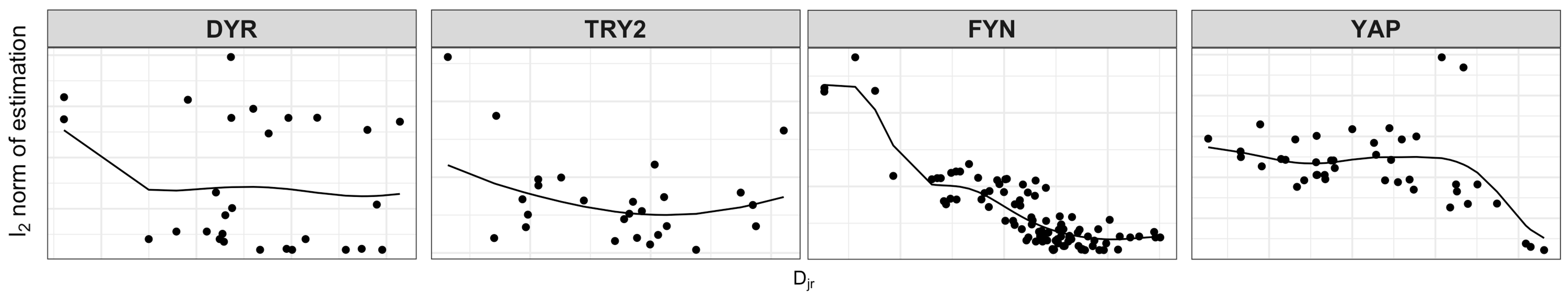}
\caption{The distances $D_{jr}$ between sites $j$ and $r$ versus $\|\widehat{\bgamma}^{(g)}_{j(r)}\|_2$ with the fitted trend.}
\label{fig:motivation}
\end{figure}

To determine the form of $\mathcal{K}(\cdot)$ from data, we draw inspiration from the adaptive group Lasso \citep{huang2008adaptive}. We consider four protein families: DYR, Trypsin-2 (TRY2), Tyrosine-protein kinase Fyn (FYN), and Yes-associated protein (YAP), and estimate $\widehat{\bgamma}^{(g)}_{j(r)}$ using multinomial regression with group Lasso for all site pairs $(j, r)$. Next, we examine the relationship between $\|\widehat{\bgamma}^{(g)}_{j(r)}\|_2$ and $D_{jr}$. As shown in Figure \ref{fig:motivation}, there is a consistent trend across the protein families under considerations, with greater magnitude of dependency parameters observed at spatially closer sites. This aligns with the biological intuition that direct couplings within sequences are expected to be stronger at spatially closer sites. It suggests that $\mathcal{K}(\cdot)$ should assign larger group weights to more distant sites to encourage greater sparsity in their dependency parameters. Based on this observation, we consider 
\begin{equation}
\mathcal{K}(D_{jr})=1-\exp(-D_{jr}^2/\mathrm{MS}_j)\label{KD},
\end{equation} 
where $\mathrm{MS}_j=(d-1)^{-1}\sum_{r\neq j}[D_{jr}-(d-1)^{-1}\sum_{r'\neq j}D_{jr'}]^2$ is
chosen to normalize the distances between site $j$ and other sites \citep{li2002kernel, wang2009gaussian}. In Section \ref{sec:simulation}, we also explore alternative forms of $\mathcal{K}(\cdot)$ to demonstrate the robustness of the choice in \eqref{KD}.

\begin{remark}
When group weights are considered, the theoretical results in Section \ref{sec:theory} can be easily extended to cases where the weights are fixed or derived from an independent dataset, such as the protein's structural data, by adjusting $\lambda_{\sharp}$ of \eqref{eq:def-lambda-sharp} to $\lambda_{\sharp}/w_{\min}$, where $w_{\min}=\min_{r\ne j, w_{jr}\ne 0} w_{jr}$. As $w_{\min}$ is independent of the sample size of the training dataset, this adjustment has minimal impact on the convergence rate.
\end{remark}

\section{Protein Mutation Analysis and Fitness Landscape}\label{sec: real}
To demonstrate our method for mutation analysis, we apply it to MSA datasets from 12 protein families, each accompanied by a representative protein structure and experimentally evaluated mutation effects.

In MSA data analysis, sequence sampling bias commonly arises from closely related species, leading to an overrepresentation of protein sequences with high similarities. This imbalance in taxonomic diversity skews mutation rate sampling and introduces spurious correlation signals \citep{hopf2017mutation,marks2011protein}. Following \cite{morcos2011direct}, we used sample weights $\bomega=(\omega_1,...,\omega_n)^\T$ to account for redundancies within a protein family by applying a threshold to the normalized Hamming distances between sequences, where $\omega_i = \{\sum_{i'\neq i}\indicator{D_{\mathrm{HM}}(\bx_{-j}^{(i)},\bx_{-j}^{(i')})<0.2}\}^{-1}$ and $D_{\mathrm{HM}}$ is the normalized Hamming distance. In practice, sample weights can be computed from the same data or estimated from an independent source. Although our theoretical results assume equal sample weights, they readily extend to cases with fixed or independently estimated sample weights. 

\subsection{Predicted energy changes versus experimental fitness} \label{subsec: RealData}
As discussed in Section \ref{sec:method}, our method evaluates energy changes, reflecting the relative favorability of site-specific mutations. To validate our method, we compute the Spearman correlation between our estimated energy changes and experimentally determined mutation fitness. As shown in Section \ref{sec:simulation} later, our method is relatively robust when $\mathcal{K}(D_{jr})$ is of the form in \eqref{KD}, reflecting the biological intuition that direct couplings between sites tend to decrease as inter-site distances increase. Thus, we adopt group weights \eqref{eq:def-group-weight} with $\mathcal{K}(D_{jr})$ of the form in \eqref{KD} for analysis. We compare our results with predictions from other methods, including the widely-used EVM, which does not account for residue structural information or group sparsity. We also consider three sparse group Lasso estimators: no weights (SGL) with all $w_{jr}=1$ in \eqref{eq:penalty_weighted}, adaptive weights, and refitted weights. Both estimators with adaptive and refitted weights start with an initial estimate $\widehat{\bgamma}^{(g)}$ from group Lasso with $w_{jr}=(\sqrt{d_r/n}+\sqrt{2\log(d-1)/n})$. For refitted weights, we use the generalized additive model in \texttt{R} package \texttt{mgcv} to fit $D_{jr}$ and $\|\widehat{\bgamma}_{j(r)}^{(g)}\|_2$, yielding $\|\widehat{\bgamma}_{j(r)}^{(g)}\|_2 = \widehat{f}(D_{jr})$, and then re-run the node-wise multinomial regression with a sparse group Lasso penalty and group weights $w_{jr}=[\widehat{f}(D_{jr})]^{-1}$. While refitted weights incorporate structural information between sites, they do not specifically model the decay of direct couplings with distance. For adaptive weights, we re-run the node-wise multinomial regression with group weights $w_{jr}=\|\widehat{\bgamma}_{j(r)}^{(g)}\|_2^{-1}$, which are adaptive \citep{huang2008adaptive} yet do not account for structural information.

\begin{table}[ht]
\centering
\caption{Spearman correlations between estimated energy change and experimental mutation fitness.}
\label{table:spearman}
\begin{tabular}{lccc|l|ccccc}
\toprule
 Protein  &  Sequences  &  Sites  & Exp. &  Mutation & Our & \multirow{2}{*}{EVM} & Refitted  & Adaptive & \multirow{2}{*}{SGL}\\  
Family& Number $(n)$ & $(d)$ &Data &Feature& Method &&Weights&Weights& \\
\midrule
 BLAT  & 8403 & 263 & 4611 &$T_m$ & \textbf{0.65} & 0.57 & 0.35 &0.42 & 0.37 \\ \midrule
 DLG4  &   102410  &  101 & 1577  & CRIPT & \textbf{0.55} & 0.54 & 0.41 & 0.37  & 0.39 \\ \midrule
 DYR & 8494 & 158 & 16 & abundance 37 & \textbf{0.86} & 0.75 & 0.81 & 0.83 &0.73 \\ \midrule
 FYN & 115571 & 66 & 42 & $T_m$ & 0.70 & 0.63 & 0.66 & \textbf{0.73} & 0.62 \\ \midrule
 GAL4 & 17521  &  75 & 1196  &  SEL & \textbf{0.64}  & 0.59 & 0.52& 0.60 &0.48  \\ \midrule
 HSP82  & 15329 & 240 &  4323 & SEL & \textbf{0.57} & 0.49  & 0.24 & 0.30 &0.33\\ \midrule
 KKA2& 12861 & 264 & 4385 &Kan$_{1:8}$ & 0.62 & 0.49 & 0.39&  \textbf{0.67} &0.42\\ \midrule
PYP  &  124287 & 125 & 125 & $\Delta G_U$  & \textbf{0.57} & 0.52 & 0.40 &0.35  &0.42\\ \midrule
YAP1 &   40302  &  36 & 363 &linear & \textbf{0.63}  &   0.44 & 0.58 & 0.57 &0.49 \\ \midrule
MTH3 &  14115  & 330 & 1957 & $W_{rel}$  & \textbf{0.52}  &  0.51 & 0.16 & 0.38  & 0.46\\ \midrule
TRY2 &  47913 & 223 & 14 & $\log(k_{cat}/K_m)$ & \textbf{0.14} &  -0.13   &  0.13 &  \textbf{0.14} & 0.10 \\ \midrule
UBE4B&  9172 & 104 & 900 & $\log_2$ ratio  &   \textbf{0.47} & 0.42 & 0.19 & 0.45 & 0.32\\
\bottomrule
\end{tabular} 
\end{table}

Table \ref{table:spearman} summarizes mutagenesis experiments of $12$ protein families. These protein families encompass a diverse array of biological functions, including enzymatic activity in antibiotic resistance (e.g., Beta-lactamase TEM (BLAT), Aminoglycoside 3'-phosphotransferase (KKA2)), digestion (e.g., TRY2), molecular chaperoning under stress (e.g., yeast ortholog of heat shock protein 90 (HSP82)), and cellular signaling and regulation (e.g., Disks large homolog 4 (DLG4), FYN, Galactose-responsive transcription factor 4 (GAL4), and Ubiquitination Factor E4B (UBE4B)). The column "Exp. Data" indicates the number of experiments conducted, each introducing a single mutation at one site on the wild-type sequence. The "Mutation Feature" column lists measures used to quantify mutation fitness, with various metrics applied in these studies. We denote $T_m$ as the denaturation midpoint temperature, where $50\%$ of proteins are folded, analogous to thermodynamic stability ($\Delta G_U$). {\it CRIPT} refers to cysteine-rich interactor \citep{mclaughlin2012spatial}, while {\it abundance 37} indicates intracellular abundance at $37^{\circ} C$ \citep{bershtein2012soluble}. {\it SEL} represents functional selection coefficients \citep{kitzman2015massively}. {\it Linear} and {\it $\log_2$ ratio} are different functions performing on the enrichment of variants \citep{araya2012fundamental,starita2013activity}, and $W_{rel}$ denotes relative fitness effects \citep{rockah2015systematic}. Finally, $k_{cat}/K_m$ represents the conversion rate at minimal substrate concentration, and {\it Kan$_{1:8}$} refers to kanamycin substrate with aminoglycosides at 1:8 dilutions \citep{melnikov2014comprehensive}.
 
Table \ref{table:spearman} shows that the estimated energy changes from our method exhibit a strong correlation with experimental mutation fitness across all protein families. Particularly, our method outperforms all other methods for ten out of all twelve families, with slightly lower performance than the adaptive weights method in the remaining two. This confirms the advantages of incorporating structural information and group sparsity in predicting mutation fitness. 

\subsection{Mutation analysis for the Dihydrofolate reductase protein}

\begin{figure}[ht]
\centering
\includegraphics[width=14cm]{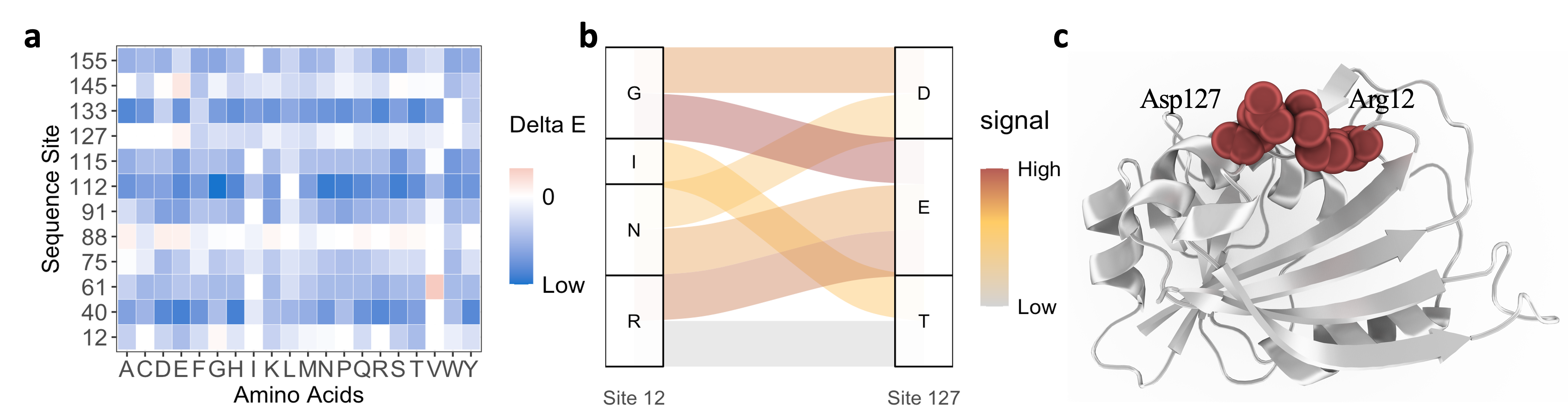}
\caption{Predicted mutation fitness of DYR. (a) Landscape of estimated energy changes for different amino acids occurring at various sites. (b) Sankey plot showing amino acid-wise dependencies between sites 12 and 127. (c) Two coevolved residues, sites 12 (Arg12, right) and 127 (Asp127, left), forming a contact in the wild-type protein structure (DYR in E.coli). The two contacting residues are highlighted in red and shown in the sphere representation.}
\label{fig:DYR}
\end{figure}

We showcase our method for mutation analysis on two protein families: DYR in this section and Postsynaptic density protein 95 (PSD95) in Sec \ref{subsec:PSD95}. Dysregulation of DYR activity is associated with various diseases and studying mutation patterns in the DYR family has broader implications for health and disease treatment \citep{baccanari1981effect,schweitzer1990dihydrofolate}. Figure \ref{fig:DYR}(a) presents the {\it fitness landscape}, where each block represents $\widehat{\Delta E}_{j,k}$, as defined in \eqref{eq:deltaE}, for the mutant of amino acid $k$ from the wild-type at site $j$. The $y$-axis corresponds to selected sites, and the $x$-axis represents the $20$ amino acid types. The color gradient (blue to white to red) reflects increasing $\widehat{\Delta E}_{j,k}$, with higher values (red) indicating favorable mutations and lower values (blue) indicating unfavorable ones. Wild-type amino acids at each site are shown in white. This landscape helps identify mutations favored by evolution. In this landscape, sites such as 40, 112, 115, and 133, shown in blue, represent highly conserved sites where most amino acid changes are unfavorable. Conversely, sites such as 12, 88, 127, and 145, shown in lighter red tones, exhibit greater tolerance to mutations. 

Amino acid variations at coevolved site pairs exhibit strong mutation dependence, indicating constraints on changes at these sites and they often interact within close spatial proximity in the protein structure. Our approach calculates amino acid-wise dependencies between coevolved sites, which offers finer resolution of interactions between different amino acid types. Figure \ref{fig:DYR}(b) presents a Sankey plot illustrating amino acid-wise dependencies between sites 12 and 127. Each side lists amino acid types observed at the MSA data, with connections between sites representing $\gamma_{jr,kl}+\theta_{jk}+\theta_{rl}$. Dark orange connections indicate a preference for co-occurrence of two amino acids, whereas lighter-colored connections indicate repulsion. 

In Figure \ref{fig:DYR}(b), we show two pairs of highly dependent amino acid types: (1) polar amino acids Asparagine (N) and Arginine (R) at site 12 with polar amino acids Aspartate (D) and Glutamate (E) at site 127, and (2) hydrophobic amino acids Glycine (G) and Isoleucine (I) at site 12 with polar amino acids Aspartate (D), Glutamate (E), and Threonine (T) at site 127. The amino acid composition at site 127 suggests conservation of polar amino acids, while site 12 shows greater tolerance for amino acids with different properties. The interaction between polar amino acids indicates potential charge compensation is preferred at those sites. In Figure \ref{fig:DYR}(c), we show that the two residues corresponding to sites 12 and 127 are in close contact on the protein structure of DYR in E.coli. A covalent bond may form between the negatively charged Aspartate (D, left) and the positively charged Arginine (R, right).

\subsection{Mutation analysis for the Postsynaptic density protein}\label{subsec:PSD95}

\begin{figure}[h]
\centering
\includegraphics[width=14cm]{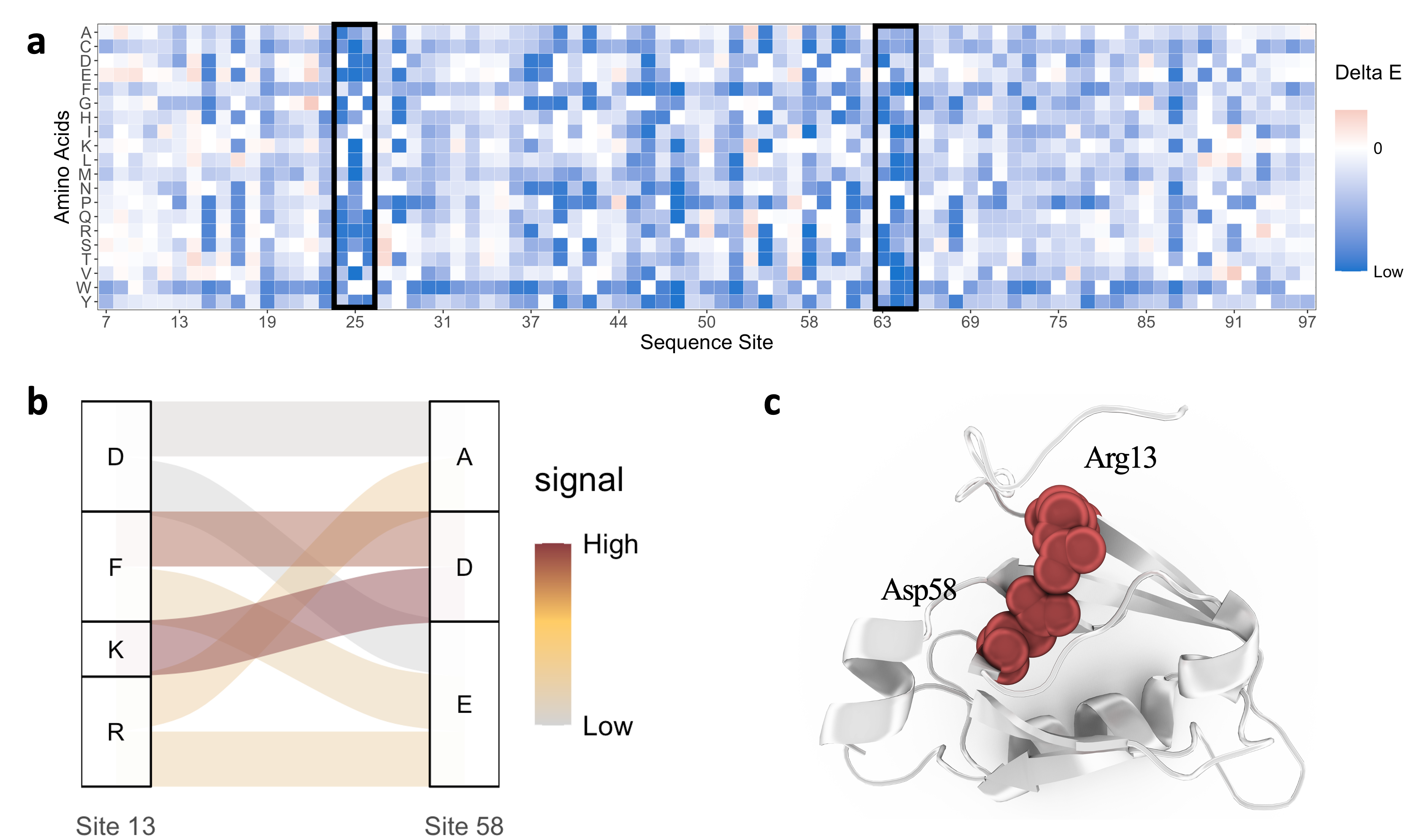}
\caption{Predicted mutation fitness of DLG4. (a) Landscape of estimated energy change for different amino acids at each sites. The framed regions exhibit low mutation fitness. (b) The Sankey plot showing amino acid-wise dependence between sites 13 and 58. (c) Protein structure with a contact formed between sites 13 (Arg13, top) and 58 (Asp 58, bottom). The two contacting residues are highlighted in red and shown in the sphere representation.}
\label{fig:DLG4}
\end{figure}

PSD95, also known as Disks large homolog 4 (DLG4), is a postsynaptic scaffolding protein involved in synaptogenesis and synaptic plasticity \citep{prange2004balance,gao2023transcriptome}. The mutation fitness landscape of the DLG4 family generated by our method is presented in Figure \ref{fig:DLG4}(a). 

Figure \ref{fig:DLG4}(a) highlights framed regions covering sites $24$-$26$ and $63$-$65$, with low average mutation fitness values (mostly blue), suggesting weak tolerance to mutations at these sites. These regions are more conserved in the MSA data, often indicating functional or structural importance. Figure \ref{fig:DLG4}(b) shows amino acid-wise dependencies between two strongly dependent sites, 13 and 58. Similar to Figure \ref{fig:DYR}(b), each side of the Sankey plot in Figure \ref{fig:DLG4}(b) lists amino acid types observed at the MSA data, with connections indicating the strength of the dependency between amino acid types. 

We consider a pair of coevolved sites, 13 and 58, to demonstrate the mutation effect predicted by our method. The DLG4 protein structure from Rat is shown in Figure \ref{fig:DLG4}(c), where the interaction between the two corresponding residues is confirmed by their close proximity. A covalent bond may form between a positively charged Arginine (R, top) and a negatively charged Aspartate (D, bottom), with their side chains pointing towards each other. Our method identifies strong mutation dependence between these two sites, with amino acid-wise dependencies shown in Figure \ref{fig:DLG4}(b). Strong connection between Lysine (K) at site 13 and Aspartate (D) at site 58, as well as Arginine (R) at site 13 and Glutamate (E) at site 58, is shown in dark orange connections, suggesting a clear pattern of charge compensation. In contrast, the co-occurrence of Aspartate (D) at site 13 and Glutamate (E) at site 58 in the sequence is unfavorable due to the same charges, as shown by the gray connection. Additionally, strong dependence is observed between Phenylalanine (F) and Aspartate (D) and between Alanine (A) and Arginine (R). While these pairs do not fit a clear property compensation pattern, they could be explained by higher-order dependencies involving multiple coevolved sites or false signals arising from misalignments in MSA data.

\section{Evidences from Numerical Experiments}\label{sec:simulation}
In this section, we evaluate our proposed method numerically and compare it with following: node-wise Lasso with $\lambda_g=0$ in \eqref{eq:penalty_weighted}, node-wise sparse group Lasso with $w_{jr}=1$ for all $j,r$ in \eqref{eq:penalty_weighted}, and node-wise ridge regression as implemented in EVM \citep{hopf2017mutation}.

\subsection{Settings and implementations}\label{sec:simulation-setting}

For all numerical experiments, we generate $n$ independent $d$-dimensional sequences. Data generation for the Potts model is non-trivial due to the computational challenges of the large state space of $\bz$, even for modest $K$ and $d$ \citep{izenman2021sampling}. To address this, we use a Gibbs sampler based on the conditional probability in \eqref{conditional_probability} to sequentially sample each site. Further details on the data generation are provided in Section C.1 of the Supplementary Material. 

The independent entries of model parameters $\btheta_{d\times K}$ are generated from $\mathrm{Unif}(0,2)$. To introduce structural information among sites, we generate a distance matrix $\bD$ with independent symmetric entries $D_{jr}\sim 40\mathrm{Beta}(2,2)$ for $j<r$, where $D_{jr}$ provides the distance between sites $j$ and $r$. 

We also generate a binary adjacency matrix $\bA$ with independent entries $A_{jr}\sim \mathrm{Ber}(p_{jr})$, where $A_{jr}=0$ indicates $\bgamma_{j(r)}=0$. Here, $p_{jr}\in(0,1)$ controls group-wise sparsity $s_g$. For element-wise sparsity, we set $\gamma_{jr,kl}=0$ for $6\leq k,l\leq K$ and each $j,r$. We consider two settings of coefficients: (M1) where the magnitude of coefficients, i.e., the signal strength between sites, is related to their distance, and (M2) where the connection probability between sites is distance-dependent.
\begin{itemize}
\item[(M1)] Set $p_{jr} = \log d/(2d)$, making the site-wise connections sparse \citep{bollobas2011sparse}. If $A_{jr}=1$, we set $\gamma_{jr,kl}= \exp(-D_{jr}^2/\mathrm{MS}_j) u_{jr,kl}$ as nonzero entries for all $1\le k,l \le 5$, where $u_{jr,kl}$ is independently sampled from $\mathrm{Unif}([-2,-0.5]\cup [0.5,2])$, and $\mathrm{MS}_j$ is defined in \eqref{KD}.
    \item[(M2)] The nonzero entries $\gamma_{jr,kl}$ are independently sampled from $\mathrm{Unif}([-2,-0.5]\cup [0.5,2])$, with $p_{jr} = \tau\exp(-D_{jr}^2/\mathrm{MS}_j) [\sum_{r'}\exp(-D_{jr'}^2/\mathrm{MS}_j)]^{-1}$, where $\mathrm{MS}_j$ is defined as in (M1). To control the group sparsity $s_g$, $\tau = 3$ is used for $d=25$ and $\tau = 1.5$ for $d=50$.
\end{itemize}

Tuning parameters for all methods are selected via 5-fold cross-validation. For our method and node-wise sparse group Lasso, which involve two tuning parameters $\lambda_g$ and $\lambda$, we search over the grid $\{(i2^{j}, (1-i)2^{j}): i\in I; j\in J\}$ with $I=\{0,0.1, 0.2,...,1\}$ and $J=\{-5,-4,-3,-2,-1,-0.5,0,0.5,1,2\}$ to reduce computational cost. If the simulated data are highly imbalanced such that frequency of $z_{jk_0}=1$ falls below $10$ for some $j\in[d]$ and $k_0\in[K]$, the corresponding observations are excluded, and we set $\widehat{\bgamma}_{i\bullet,k_0\bullet}=0$ by default to ensure valid cross-validation.

We also explore two ways for constructing group weights $w_{jr}$ of the form \eqref{eq:def-group-weight} using different $\mathcal{K}(D_{jr})$ to examine how the choice of $\mathcal{K}(D_{jr})$ affects coefficient estimation. In (N1), $\mathcal{K}(D_{jr})$ matches the form used in the data generating process, while in (N2), it differs but retains a similar trend. 
\begin{itemize}
    \item[(N1)] Set $\mathcal{K}(D_{jr})=1-\exp(-D_{jr}^2/\mathrm{MS}_j)$, consistent with how $D_{jr}$ is used for generating coefficients in (M1) and (M2). 
    \item[(N2)] Set $\mathcal{K}(D_{jr})=\exp(D_{jr})[1+\exp(D_{jr})]^{-1}$, commonly used to generate adjacency matrices in the Erd\H{o}s-R{\'e}nyi model in network analysis.
\end{itemize}

\subsection{Results}
To assess the estimation accuracy of a method, we use the mean squared estimation error (MSE), i.e., $\sum_{j=1}^d\|\bgamma_j-\Tilde{\bgamma}_j\|_2^2$. For selection accuracy, we consider the true positive rate (TPR) and the false discovery rate (FDR) for identifying nonzero $\gamma_{jr,kl}$ and groups $\bgamma_{j(r)}$. To clarify, TPR is defined as TP/(TP+FN), where TP and FN are the numbers of correctly and incorrectly estimated nonzero parameters, and FDR is FP/(FP+TP), where FP is the number of incorrectly estimated nonzero parameters. 
The TPR and FDR for identifying nonzero groups are defined similarly and denoted as TPR$_{\mathrm{g}}$ and FDR$_{\mathrm{g}}$. We evaluate our model and competitors under (M1) and (M2) with $n=1000, 2000, 4000$, $d=25, 50$, and $K=20$. All measures are computed based on $100$ independent Monte Carlo replicates.

Table \ref{table:1} reports the mean squared error (MSE) along with entry-wise and group-wise TPRs and FDRs for the (M1) setting. The results show that our method outperforms competitors in both estimation and selection accuracy across varying sample sizes $n$ and numbers of sites $d$ when signal strengths depend on distances. Similarly, as shown in Table C1 in the Supplementary Material, our method also excels under (M2) when site connection probabilities are determined by their distances. These findings confirm the effectiveness of our method for accurately estimating coefficients when signal strengths or inter-site connections are physically distance-dependent. By incorporating distance information into parameter estimation, our method provides reliable estimates, even when the chosen $\mathcal{K}(D_{jr})$ only approximates the desired trend, as in (N2). This robustness supports the form in \eqref{KD} for physical distance-based group weights in practice. Additionally, estimation errors decrease and selection accuracy improves as $n$ increases or $d$ decreases, consistent with Theorem \ref{thm:est-err-rates}. Owing to space constraints, detailed results for (M2) and more settings with greater $d$ and $n$, aligned with the real data in Section \ref{sec: real}, are presented in Section C.2 of the Supplementary Material.
\begin{table}[!ht]
\centering
\caption{Results in (M1) with $K=20$}
\label{table:1}
\begin{tabular}{c|c|l|c|ccccc}
\toprule 
$d$ & $\sum_{j<r}A_{jr}$ &Methods & $n$  &  MSE &  TPR & FDR & TPR$_{\mathrm{g}}$ & FDR$_{\mathrm{g}}$  \\  
\midrule
\multirow{15}{*}{25} &\multirow{15}{*}{25} &Our method with (N1) &\multirow{5}{*}{1000}&  24.595 & 0.798 &  0.054 & 0.960 &0.250 \\
& &Our method with (N2) & &  30.332 & 0.747 &0.084 & 0.940 & 0.242\\
& &Lasso  & & 47.192 & 0.694 & 0.392 & 0.860 & 0.403\\
& &Sparse Group Lasso & & 38.513 & 0.731 & 0.120& 0.920 & 0.270\\
& &Ridge & & 84.050 & -- & --& -- & -- \\
\cmidrule{3-9}
& & Our method with (N1) &\multirow{5}{*}{2000}
& 17.368 & 0.842 & 0.051 & 0.980 & 0.197\\
& &Our method with (N2) & & 25.357 &0.790 & 0.080& 0.980 & 0.210\\
& &Lasso  & & 36.148 &0.724 &0.352  & 0.900 & 0.365 \\
& &Sparse Group Lasso & &  29.429 & 0.757 & 0.116 & 0.940& 0.254 \\
& &Ridge & & 60.954  & -- & -- & --& --  \\
\cmidrule{3-9}
& & Our method with (N1) &\multirow{5}{*}{4000}
& 14.934& 0.871 &0.043&  1.000 & 0.138 \\
& &Our method with (N2) & &18.172 & 0.813& 0.069  & 1.000 &0.153\\
& &Lasso  & &  27.637& 0.788 & 0.314 & 0.960 & 0.308 \\
& &Sparse Group Lasso & &   23.080 & 0.810  & 0.085 &1.000 & 0.242 \\
& &Ridge &  &  48.695 & -- & -- & --& --  \\
\midrule
\multirow{15}{*}{50} &\multirow{15}{*}{78} &Our method with (N1) &\multirow{5}{*}{1000}& 67.819 & 0.758 & 0.077  & 0.923 &0.242 \\
& &Our method with (N2) & & 73.855 & 0.712 & 0.105 & 0.904 & 0.265\\
& &Lasso  & &94.631 & 0.633 & 0.374 & 0.852 & 0.355  \\
& &Sparse Group Lasso & & 80.518 & 0.686 & 0.177 & 0.846 & 0.284 \\
& &Ridge & & 127.514  & -- & -- & --& -- \\
\cmidrule{3-9}
& &Our method with (N1) &\multirow{5}{*}{2000}& 50.145 & 0.803  & 0.073  &0.944 & 0.216   \\
& &Our method with (N2) & &57.254 & 0.762 & 0.093 &0.936& 0.239 \\
& &Lasso  & & 73.953 & 0.677& 0.293  & 0.878  & 0.327    \\
& &Sparse Group Lasso & &  62.148 & 0.724   & 0.132 &0.913  &0.253  \\
& &Ridge & & 88.367  & -- & -- & -- & -- \\
\cmidrule{3-9}
& &Our method with (N1) &\multirow{5}{*}{4000} & 41.512 & 0.849 & 0.064  & 1.000 &  0.194 \\
& &Our method with (N2) & & 44.416 & 0.810 & 0.112 & 1.000 & 0.218\\
& &Lasso  & & 51.794 & 0.720 & 0.278 & 0.926 & 0.275 \\
& &Sparse Group Lasso & & 48.268 & 0.783 & 0.112 & 1.000 & 0.243 \\
& &Ridge &  &  72.620 & -- & -- & --& --\\
\bottomrule
\end{tabular}
\end{table}

\section{Conclusions with Discussion}\label{sec:dis}
This paper studies the long-standing challenge of predicting protein mutation fitness by using the high-dimensional Potts model. To estimate model parameters, we adopt node-wise multinomial regression with a sparse group Lasso penalty, capturing both site-wise and element-wise sparsity in protein sequences. Our method balances selecting significant site pairs with identifying key amino acid-level interactions, and incorporating protein structural data via group weights enhances the estimation accuracy of parameters. Theoretically, we extend recent oracle inequalities for high-dimensional linear models \citep{bellec2018slope} to derive error bounds for our estimator, underscoring its validity. These error bounds match the minimax lower bound for double sparse linear regression, up to a factor specific to the multinomial structure of the Potts model. Experimentally, our method outperforms existing competitors, particularly the widely-used EVM in predicting mutation fitness, validated by high-throughput mutagenesis experiments across multiple protein families. This suggests the potential of our method to advance the understanding of protein functionality in an evolutionary context. 

\textit{Limitations and Challenges.} While our method shows promising results, several areas warrant further exploration. Our method assumes that spatial distances between sites is the primary factor that determines the strength of direct couplings. Including additional factors, such as solubility and solvent accessibility, could enhance predictions and merit investigations. Furthermore, our approach focuses on pairwise effects, but higher-order interactions involving three or more sites may also significantly impact protein functionality. Efficient algorithms to address these higher-order dependencies remain a key challenge for future work.

While parallel computing with node-wise multinomial regressions mitigates some of the demanding computational costs of our method, the high dimensionality of the categorical design matrix, caused by many sites and categories, makes the algorithm slow to converge. Advances in optimization techniques could help address this issue \citep{qi2024non}. Also, incorporating different structural information into data generation from the Potts model remains challenging. Further work is needed to develop procedures that better replicate real MSA data, akin to scDesign3 \citep{song2024scdesign3} for single-cell RNA-sequencing data.

Our theoretical guarantees focus on estimators from multinomial regression. While node-wise multinomial regressions provide a solid foundation for estimating the Potts model, the lower bound for estimation in multinomial regression over the double sparse parameter class remains unknown. Consequently, it is unclear whether the derived error bounds fully capture the complexities inherent in multinomial regressions, which is left for future studies. 

\textit{Extensions and Future Directions.}
While focused on protein mutations, the proposed method has broader applications in computational biology, such as detecting gene regulatory networks or metabolic pathways, where hierarchical and spatial dependencies are prevalent. The model can also be refined to account for dynamic aspects of protein interactions, incorporating temporal changes in coevolutionary patterns. Our theoretical results can readily extend to accommodate fixed or independently estimated sample weights (for sequence sampling bias) and group weights (for structural information integration) in the penalty. However, in practice, these weights may be derived from the same dataset, a challenge that could be addressed through dedicated analysis or data splitting techniques. Moreover, while this work emphasizes the estimation of the Potts model, statistical inference for the Potts model remains largely unexplored yet holds significant values for understanding uncertainties in predicting protein mutation fitness. We plan to address these directions in future research.

\noindent
{\bf Data availability.} MSA data of protein families considered in this paper and the reproducible code for simulation studies can be found at GitHub (\url{https://github.com/BingyingD/Potts-Spatial-Protein-Mutation}).

\bibliographystyle{apalike}

\end{document}